\UseRawInputEncoding

\documentclass[aps, preprint, groupedaddress, floatfix]{revtex4}

\usepackage{epsfig}
\usepackage{epstopdf}
\usepackage{graphicx}
\usepackage{amsmath}
\usepackage{color}
\usepackage{ragged2e}
\usepackage{makecell}
\usepackage{array}
\usepackage{verbatim}

\def\[{\begin{equation}}
\def\]{\end{equation}}
\def\bma{\begin{pmatrix}}
\def\ema{\end{pmatrix}}

\begin{document}

\title{Emergent topological states via digital (001) oxide superlattices}

\author{Zhiwei Liu$^{1,2,3,\#}$, Hongquan Liu$^{2,\#}$, Jiaji Ma$^2$, Xiaoxuan Wang$^2$, Gang Li$^{4,5,*}$ and Hanghui Chen$^{2,3,6,*}$}

\affiliation{$^1$Key Laboratory of Polar Materials and Devices, Ministry of Education, East China Normal University, Shanghai 200241, China.
\\
  $^2$NYU-ECNU Institute of Physics, NYU Shanghai, Shanghai 200122, China\\
  $^3$Department of Electronic Science, East China Normal University, Shanghai 200241, China.\\
  $^4$School of Physical Science and Technology, ShanghaiTech University, Shanghai 201210, China \\
  $^5$ShanghaiTech Laboratory for Topological Physics, ShanghaiTech University, Shanghai 201210, China \\
  $^6$Department of Physics, New York University, New York 10012, USA \\
  $^*$Email: ligang@shanghaitech.edu.cn;  $^*$Email: hanghui.chen@nyu.edu
}
\date{\today}

\begin{abstract}
Oxide heterostructures exhibit many intriguing properties. Here we
provide design principles for inducing multiple topological states in
(001) ($AM$O$_3$)$_1$/($AM'$O$_3$)$_1$ oxide superlattices. Aided by
first-principles calculations and model analysis, we show that a
(Sr$M$O3)$_1$/(Sr$M'$O$_3$)$_1$ superlattice ($M$ = Nb, Ta and $M'$ = Rh, Ir) is a
strong topological insulator with $Z_2$ index (1;001). More remarkably, a
(SrMoO3)$_1$/(SrIrO3)$_1$ superlattice exhibits multiple coexisting
topological insulator (TI) and topological Dirac semi-metal (TDS)
states. The TDS state has a pair of type-II Dirac points near the
Fermi level and symmetry-protected Dirac node lines. The surface TDS
Dirac cone is sandwiched by two surface TI Dirac cones in the
energy-momentum space. The non-trivial topological properties arise from the
band inversion between $d$ orbitals of two dissimilar transition metal
atoms and a particular parity property of (001) superlattice
geometry. Our work demonstrates how to induce nontrivial topological
states in (001) perovskite oxide heterostructures by rational design.
\end{abstract}

\maketitle

\section*{Introduction}

Complex oxides exhibit a wide range of intriguing phenomena
such as Mott insulators~\cite{10.1103/PhysRevB.44.943,10.1088/0370-1298/62/7/303,10.1103/RevModPhys.40.677}, long-range charge/spin/orbital orders~\cite{10.1103/RevModPhys.70.1039,10.1146/annurev-conmatphys-062910-140445,10.1126/science.288.5465.462}, multiferroics~\cite{10.1038/nmat2373,10.1103/PhysRevLett.93.177402,10.1103/PhysRevB.67.180401,10.1038/s41563-018-0275-2,10.1038/natrevmats.2016.46,10.1038/nmat1804} and high-temperature superconductivity~\cite{10.1007/BF01303701,10.1103/RevModPhys.66.763,10.1038/375561a0,10.1038/nature14165,10.1103/RevModPhys.78.17}. On the other hand, searching for non-trivial topological states in solid-state~\cite{10.1103/RevModPhys.82.3045,10.1103/RevModPhys.83.1057}, photonic~\cite{10.1103/PhysRevLett.126.113902,10.1038/nphoton.2014.248} and acoustic~\cite{10.1038/s41563-021-00985-6,10.1038/s42254-019-0030-x} crystals has been one of the most active fields in condensed matter physics and materials science. 
In time-reversal-invariant crystalline solids, topological insulating state and topological metallic state (with Dirac points and/or Dirac node lines) have been intensively studied in narrow-gap semiconductors whose electronic structure is dominated by $s$ and $p$ orbitals~\cite{10.1103/PhysRevB.92.045108,10.1103/PhysRevLett.98.106803,10.1126/science.1245085,10.1038/nature06843,10.1126/science.1148047,10.1126/science.1133734,10.1038/nphys1270,10.1038/nphys1274,10.1038/nphys1689}. In comparison, complex oxides with characteristic $d$ orbitals have been less explored. Previous studies propose that bulk YiBiO$_3$~\cite{10.1038/srep01651} and electron doped BaBiO$_3$~\cite{10.1038/nphys2762} are candidates for topological insulators, in both of which a $s$-$p$ band inversion occurs and spin-orbit coupling opens a gap, similar to Bi$_2$Se$_3$~\cite{10.1038/nphys1270}. Crystal symmetry (in particular with non-symmorphic space group) has been exploited to search for Dirac points and Dirac node lines in non-magnetic oxides such as SrNbO$_3$~\cite{10.1103/PhysRevB.104.235121} and SrIrO$_3$~\cite{10.1103/PhysRevB.85.115105,10.1038/ncomms7593}. Recently, high-mobility and giant magnetoresistance have been observed in Dirac semi-metal CaIrO$_3$~\cite{10.1038/s41467-018-08149-y,10.1103/PhysRevLett.123.216601}. 

The advances in thin-film deposition techniques enable the synthesis of digital oxide superlattices and heterostructures, providing a different approach to inducing non-trivial topological states in complex oxides~\cite{10.1103/PhysRevB.84.201103,10.1103/PhysRevB.88.155119,10.1103/PhysRevB.90.165143}. In particular, oxide superlattices or heterostructures can furnish artificially engineered electronic structure that is lacking in bulk~\cite{10.1038/nmat3223,10.1038/s41563-021-00989-2}. An intriguing proposal is to study a bi-layer of perovskite oxide $AM$O$_3$ thin film along (111) direction~\cite{10.1038/ncomms1602,10.1063/5.0009092}, in which the transition metal atom $M$ resides on a buckled honeycomb lattice. The transition metal $d$-bands generically form a linear crossing at the high-symmetry $K$ point in the Brillouin zone and spin-orbit coupling (SOC) opens a gap at the crossing~\cite{10.1103/PhysRevLett.95.146802} and thus non-trivial topological states may emerge such as quantum spin Hall state~\cite{10.1103/PhysRevLett.95.226801,10.1103/PhysRevLett.102.256403}. When local interactions are considered, other more exotic topological states such as quantum anomalous Hall and fraction quantum Hall states are also possible~\cite{10.1103/PhysRevB.47.7312,10.1002/adma.202001460,10.1103/PhysRevLett.61.2015}. However, (111) terminations of perovskite oxide $AM$O$_3$ are polar and it is very difficult to synthesize such films with a precise control of their thickness in experiment~\cite{10.1063/5.0009092}. By constrast, (001) perovskite oxide heterostructures have been routinely synthesized~\cite{10.1126/sciadv.aay0114,10.1111/j.1551-2916.2008.02556.x}, in particular those oxides with non-polar terminations $[A^{2+}\textrm{O}^{2-}]$ and $[M^{4+}(\textrm{O}^{2-})_2]$. They can be accurately controlled on the atomic scale in a layer-by-layer manner in an oxide superlattice~\cite{10.1038/s41578-019-0095-2}.

In this work, we provide a different approach to inducing non-trivial
topological states in complex oxides via artificially designed
time-reversal-invariant (001) oxide superlattices.  Our design
principles are based on a few factors: the band inversion between $d$
orbitals of two dissimilar transition metal atoms, a particular parity
property of the (001) superlattice geometry, oxygen octahedral
rotation pattern and $d$ orbital occupancy. Following the design
principles, we use first-principles methods and model Hamiltonian
calculations to show concrete examples. We find that a
(Sr$M$O$_3$)$_1$/(Sr$M'$O$_3$)$_1$ superlattice is a strong
topological insulator (TI) with $M$ = Nb, Ta and $M'$ = Rh, Ir. The
most promising candidate is the (SrTaO$_3$)$_1$/(SrIrO$_3$)$_1$
superlattice with the largest atomic spin-orbit coupling and a direct
gap of about 20 meV.  The topological property is robust against
epitaxial strain and weak structural distortions. More interestingly,
we show that a (SrMoO$_3$)$_1$/(SrIrO$_3$)$_1$ superlattice has
multiple coexisting topological states, including topological
insulator (TI) and topological Dirac semi-metal (TDS).  Between its
highest valence band and its lowest conduction band, there exists a
pair of Dirac points (DP) close to the Fermi level as well as Dirac
node lines (DNL) that cross the Fermi level.  The DP are stabilized by
the combination of time-reversal, inversion and $C_4$ rotation
symmetries.  They are of type-II (i.e. a tilted Dirac cone) and have a
mirror Chern number of 2 (i.e. two Fermi arcs for each DP).  The
number of DP can be controlled via epitaxial strain.  The DNL are
protected by non-symmorphic space group ($\cal{SG}$) and thus are
robust against epitaxial strain.  Between the highest valence band and
the second highest valence band, as well as between the lowest
conduction band and the second lowest conduction band, a topologically
nontrivial gap is opened with a $Z_2$ index (1;001). The origin of
this TI gap is the same as that of the (SrTaO$_3$)$_1$/(SrIrO$_3$)$_1$
superlattice. As a consequence of these multiple topological states
coexisting in the (SrMoO$_3$)$_1$/(SrIrO$_3$)$_1$ superlattice, its
surface TDS Dirac cone is sandwiched by another two surface TI Dirac
cones in the energy-momentum space.  The TDS Dirac cone and TI Dirac
cones are close to the Fermi level, implying that one can induce a
TI-TDS-TI topological state transition via chemical doping or
electric-field gating.

Our work demonstrates that artificially designed heterostructures may exhibit emergent topological properties that are absent in their bulk constituents. Therefore in addition to naturally-occurring topological materials~\cite{10.1038/s41586-019-0937-5,10.1038/s41586-019-0944-6,10.1038/s41586-019-0954-4}, heterostructuring (based on some design principles) provides a different route to emerging topological phenomena.

\begin{table}
\caption{Crystal structure parameters of candidate materials, relevant
substrates and proposed oxide superlattices in this work, including crystallographic space group ($\cal{SG}$), experimental lattice constants (EXP) and DFT-calculated lattice constants (DFT).}
\label{Table 1}
\begin{tabular}{>{\centering}m{4cm} >{\centering}m{4cm} >{\centering}m{4cm} >{\centering}m{4cm}}
\hline\hline
Material & $\cal{SG}$ & EXP ($\mathrm{\AA}$) & DFT ($\mathrm{\AA}$) \tabularnewline
\hline
SrTiO$_3$ &  $Pm\bar{3}m$ (No. 221) & $a=b=c=3.903$~\cite{10.1016/j.jallcom.2009.07.045} & $a=b=c=3.903$ \tabularnewline
KTaO$_3$  &  $Pm\bar{3}m$ (No. 221) & $a=b=c=3.983$~\cite{10.1103/PhysRevB.75.184113} & $a=b=c=3.995$ \tabularnewline
SrTaO$_3$ &  $Pm\bar{3}m$ (No. 221) & NA & $a=b=c=4.029$ \tabularnewline
SrMoO$_3$ &  $Pm\bar{3}m$ (No. 221) & $a=b=c=3.976$~\cite{10.1016/j.jssc.2009.11.005} & $a=b=c=3.957$ \tabularnewline
SrIrO$_3$ &  $Pnma$ (No. 62) & \makecell{$a=5.571,b=5.601$ \tabularnewline $c=7.896$~\cite{10.1103/PhysRevB.89.214106}} & \makecell{$a=5.630,b=5.538$ \tabularnewline $c=7.863$} \tabularnewline
\makecell{(SrTaO$_3$)$_1$/(SrIrO$_3$)$_1$ \tabularnewline on SrTiO$_3$} & $P4/mbm$ 
(No. 127) & NA & \makecell{$a=b=5.520$ \tabularnewline $c=8.048$} \tabularnewline
\makecell{(SrMoO$_3$)$_1$/(SrIrO$_3$)$_1$ \tabularnewline on KTaO$_3$} & $P4/mbm$ 
(No. 127) & NA & \makecell{$a=b=5.650$ \tabularnewline $c=7.780$} \tabularnewline
\hline\hline
\end{tabular}
\end{table}

\section*{Results}

\subsection*{Design principles}

\begin{figure}
    \centering
    \includegraphics[width=0.8\textwidth]{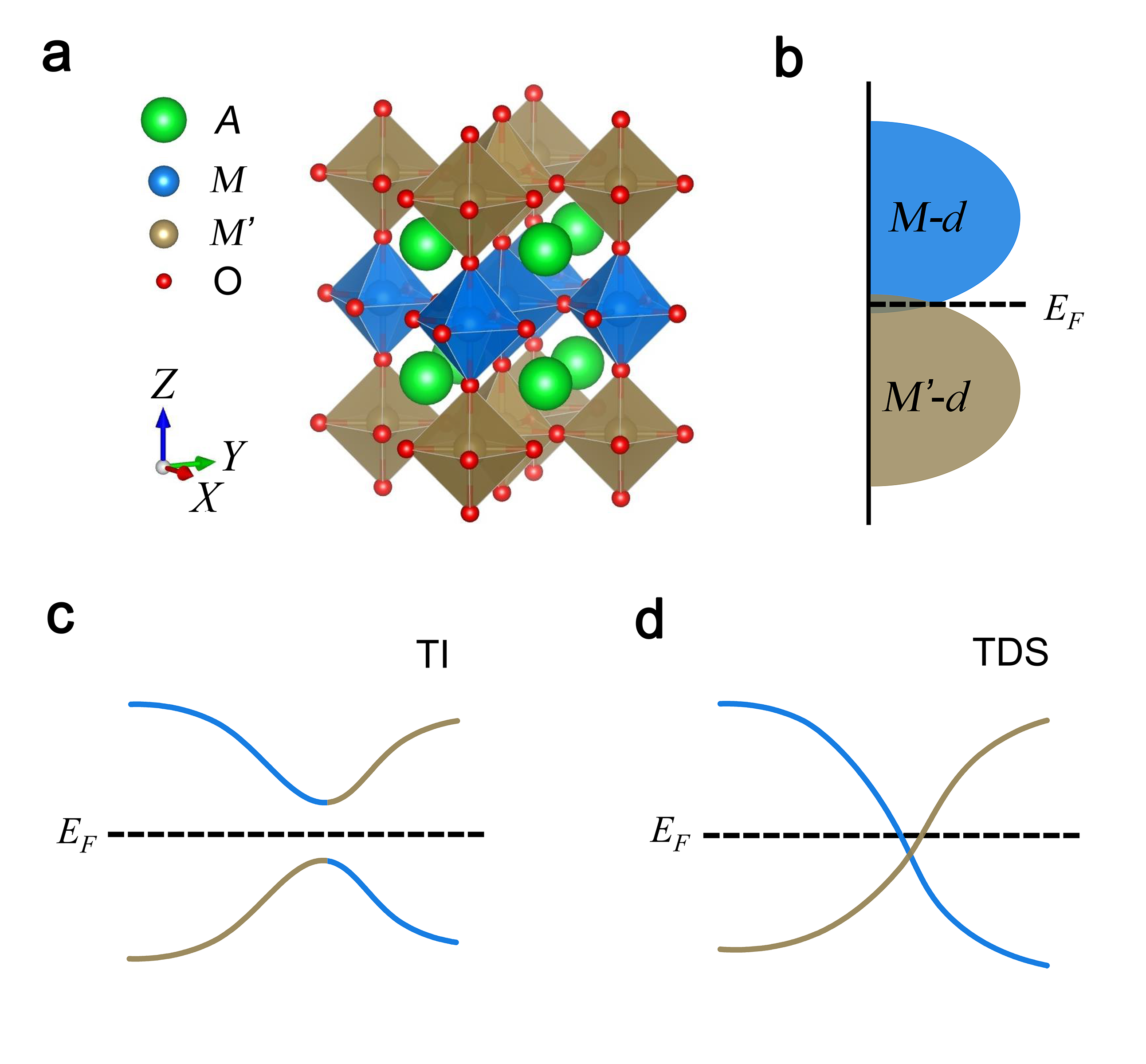}
    \caption{\textbf{Crystal structure and schematic band structure of
        (001) oxide superlattices}. \textbf{a}: The crystal structure
      of a (001) ($AM$O$_3$)$_1$/($AM'$O$_3$)$_1$ superlattice. The
      green, blue, brown and red balls represent $A$, $M$, $M'$ and O
      atoms, respectively. \textbf{b}: A schematic diagram of density
      of states of the superlattice. The blue part is the
      $M$-\textit{d} states and the brown part is the $M'$-\textit{d}
      states and the black dash line is the Fermi level. \textbf{c}
      and \textbf{d}: A schematic band structure close to the Fermi
      level where a $M$-$d$ band and a $M'$-$d$ band invert. The black
      dash line is the Fermi level. \textbf{c}: Spin-orbit coupling
      opens a full gap between the highest valence band and the lowest
      conduction band, and non-trivial topological property such as
      strong topological insulator (TI) may emerge. \textbf{d}: A
      $M$-$d$ band and a $M'$-$d$ band forms a linear crossing around
      the Fermi level in the presence of spin-orbit coupling, which
      may lead to a topological Dirac semi metal (TDS).}
    \label{fig1}
\end{figure}

Our basic idea is to construct an artificial band inversion via 
an oxide superlattice. However, different from $s$-$p$ band inversion or 
$d$-$d$ band inversion from the same transition metal atom
in the previous works~\cite{10.1038/srep01651,10.1038/nphys2762,10.1038/ncomms1602,10.1063/5.0009092,10.1103/PhysRevLett.108.140405,10.1103/PhysRevLett.112.036403},
we study $d$-$d$ band inversion 
between two different transition metal atoms in a (001) oxide superlattice. 
Panel \textbf{a} of Fig.~\ref{fig1} shows the crystal structure of a 
(001) ($AM$O$_3$)$_1$/($AM'$O$_3$)$_1$ superlattice where $M$ and $M'$ 
are transition metal atoms and $A$ is either an alkaline-earth metal or 
rare-earth metal atom. The oxygen octahedra $M$O$_6$ are corner-shared and 
they alternate along the $z$ axis.
Panel \textbf{b} shows a schematic density of states of the superlattice.
We choose $M$ to be an early transition metal atom and $M'$ to be a late 
transition metal atom. Due to electronegativity difference, the $d$ orbitals of $M$ 
have higher energies than the $d$ orbitals of $M'$~\cite{10.1103/PhysRevLett.111.116403,10.1088/1361-648X/aa6efe}. 
Given a proper combination 
of $M$ and $M'$ and their $d$ occupancy, $M$-$d$ bands and $M'$-$d$ bands 
may overlap with each other around the Fermi level. When spin-orbit coupling (SOC) 
is included, we can 
have two situations in which non-trivial topology may emerge. 
If SOC can open a full gap 
between the highest valence band and the lowest conduction band, 
a topological insulating state may emerge, given a proper interaction between 
$M$-$d$ and $M'$-$d$ states (see a schematic band structure in panel \textbf{c})~\cite{10.1126/science.1133734,10.1557/mrs.2014.216}. In this case, the (001) superlattice
geometry leads to a particular parity property of the two transition metal
$d$ orbitals, which serves as a key to understand how non-trivial topology
arises (see the next section for details).
If a linear crossing is stabilized between a $M$-$d$ band and a $M'$-$d$ 
band around the Fermi level even in the presence of SOC (see a schematic 
band structure in panel \textbf{d}), then it is 
possible to get a topological semi metallic state.

Next we discuss considerations on materials and crystal symmetries.
First we choose $A$=Sr because the formal valence of Sr is 2+. Due to
charge balance, the formal valence of $M$ and $M'$ is 4+. That leads
to non-polar terminations [Sr$^{2+}$O$^{2-}$] and
$[M^{4+}(\textrm{O}^{2-})_2]$, which are easier to control than polar
terminations in thin film
growth~\cite{10.1111/j.1551-2916.2008.02556.x}.  While the above
design principles can be extended to magnetic systems, in the current
study we focus on time-reversal-invariant systems. Therefore for $M$
and $M'$, we study second-row and third-row transition metal atoms,
which are more itinerant and less correlated than first-row transition
metal atoms~\cite{10.1038/ncomms1602}. For early transition metal
atoms such as Nb and Ta, Sr$M$O$_3$ crystallizes in a cubic structure
with no oxygen octahedral (OO) rotation ($\cal{SG}$ No. 221
$Pm\bar{3}m$, Glazer notation $a^0a^0a^0$), while for late transition
metal atoms such as Rh and Ir, Sr$M$O$_3$ crystallizes in an
orthorhombic structure with both in-plane and out-of-plane OO
rotations ($\cal{SG}$ No. 62 $Pnma$, Glazer notation $a^-b^-c^+$).
Combining a cubic $Pm\bar{3}m$ perovskite oxide with an orthorhombic
$Pnma$ perovskite oxide in a (001) oxide superlattice usually leads to
a tetragonal crystal structure ($\cal{SG}$ No. 127 $P4/mbm$, Glazer
notation $a^0a^0c^-$), which is characterized by an out-of-phase
in-plane OO rotation about the stacking
direction~\cite{10.1126/sciadv.aay0114,10.1103/PhysRevLett.108.107003}. This
crystal structure is `compromising' in that the in-plane OO rotations
are 'forced' into the cubic structure of one perovskite oxide, while
the out-of-plane OO rotations are suppressed in the orthorhombic
structure of the other perovskite oxide, resulting in only an
out-of-phase in-plane OO rotation throughout the superlattice. The
out-of-phase in-plane OO rotation in an oxide superlattice leads to
cell doubling, a $C_4$ rotation symmetry and a non-symmorphic
$\cal{SG}$ ($P4/mbm$ $\cal{SG}$ No. 127), which are crucial for Dirac
points (DP) and Dirac node lines (DNL).

Finally we discuss $d$ occupancy of $M^{4+}$ and $M'^{4+}$ ions. 
We consider two cases:
$d^1+d^5$ and $d^2+d^5$. Due to the cell doubling that is needed to 
accommodate the OO rotation, the total $d$ occupancy of the $d^1+d^5$ case 
is $(1+5)\times 2 = 12$, which can be divided by 4 implying a possible 
insulating ground state (see Supplementary 
Notes 1 for symmetry analysis). 
On the other hand, the total $d$ occupancy of 
the $d^2+d^5$ case is $(2+5)\times 2 = 14$, which is $3\times 4 + 2$. 
Given the same symmetry considerations, we must 
have a gapless system with one band half-filled 
when the total $d$ occupancy is 14.

Based on the above analysis, we use both first-principles calculations 
and tight-binding Hamiltonian to 
demonstrate that the (SrTaO$_3$)$_1$/(SrIrO$_3$)$_1$ superlattice 
is a strong topological insulator, while the (SrMoO$_3$)$_1$/(SrIrO$_3$)$_1$ 
superlattice has nontrivial topological gaps, Dirac points and 
Dirac node lines coexisting in its electronic structure, which lead 
to multiple surface Dirac cones around the Fermi level.
Table~\ref{Table 1} summarizes the candidate materials, relevant
substrates and proposed oxide superlattices in this study.

\subsection*{Strong TI state in the (SrTaO$_3$)$_1$/(SrIrO$_3$)$_1$ superlattice}

\begin{figure}
    \centering
    \includegraphics[width=0.8\textwidth]{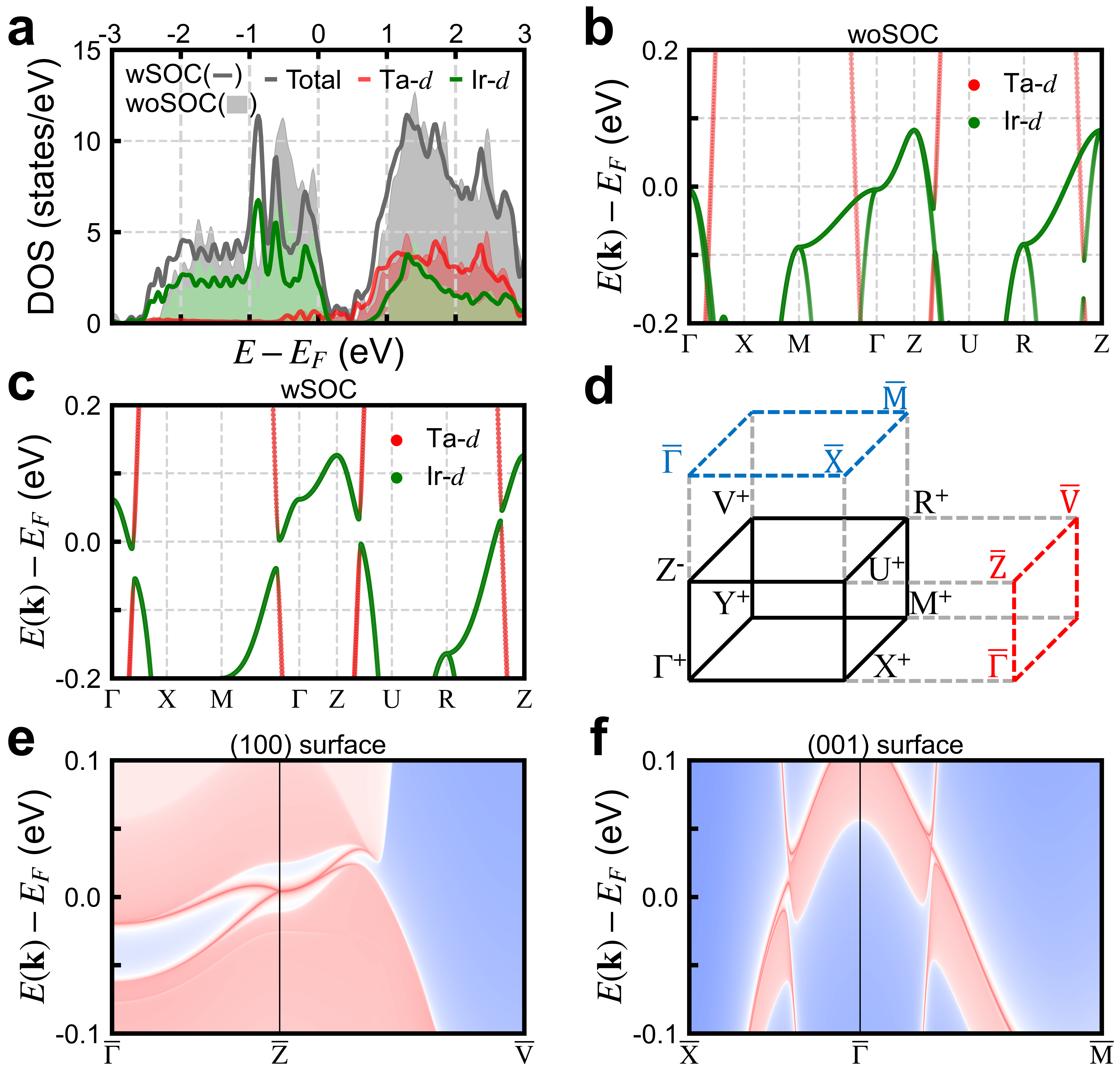}
    \caption{\textbf{Electronic and topological properties of the
        (SrTaO$_3$)$_1$/(SrIrO$_3$)$_1$ superlattice from
        first-principles calculations}. \textbf{a}: Density of states
      (DOS) of the (SrTaO$_3$)$_1$/(SrIrO$_3$)$_1$ superlattice. The
      solid curves are the DOS with spin-orbit coupling (SOC)
      included. The shades are the DOS without SOC. The gray, red and
      green curves/shades are total, Ta-$d$ projected and Ir-$d$
      projected DOS. \textbf{b}: Near Fermi level DFT band structure
      of the superlattice without SOC. The red and green symbols
      highlight the atomic projections onto Ta-\textit{d} and
      Ir-\textit{d} orbitals, respectively. The Fermi level is shifted
      to the zero energy. \textbf{c}: Near Fermi level DFT+SOC band
      structure of the superlattice. The red and green symbols have
      the same meaning as in panel \textbf{b}. A full gap is opened
      along the high-symmetry \textbf{k}-path. \textbf{d}:
      $\frac{1}{8}$ Brillouin zone of the
      (SrTaO$_3$)$_1$/(SrIrO$_3$)$_1$ superlattice. The eight TRIM
      points and their parities are explicitly shown. The red and blue
      dashed lines are $\frac{1}{4}$ surface Brillouin zone which is
      obtained by projecting the bulk Brillouin zone onto (100) and
      (001) surfaces.  \textbf{e}: Band structure of (100)
      semi-infinite slab of the (SrTaO$_3$)$_1$/(SrIrO$_3$)$_1$
      superlattice, calculated by the Green-function method. The
      surface Brillouin zone is highlighted by the red dashed lines in
      panel \textbf{d}. \textbf{f}: Band structure of (001)
      semi-infinite slab of the (SrTaO$_3$)$_1$/(SrIrO$_3$)$_1$
      superlattice, calculated by the Green function method. The
      surface Brillouin zone is highlighted by the blue dashed lines
      in panel \textbf{d}.}
    \label{fig2}
\end{figure}

We first discuss the $d^1+d^5$ case and study the (001)
(SrTaO$_3$)$_1$/(SrIrO$_3$)$_1$ superlattice grown on a SrTiO$_3$ substrate 
as a prototype. Both SrTaO$_3$ and SrIrO$_3$ are non-magnetic metals 
in bulk.
Bulk SrTaO$_3$ has a cubic structure ($\cal{SG}$ $Pm\bar{3}m$ 
No. 221), while bulk SrIrO3 crystallizes either in a monoclinic
structure of distorted 6$H$-type hexagonal perovskite 
($\cal{SG}$: $C2/c$, No. 15)~\cite{10.1038/s41535-021-00396-5} or an orthorhombic structure ($\cal{SG}$ $Pnma$ No. 62)~\cite{10.1088/0953-8984/25/12/125604,10.1063/1.4903314}. In thin films, 
the orthorhombic structure is more favored in SrIrO$_3$~\cite{10.1088/0953-8984/25/12/125604,10.1063/1.4903314,10.1103/PhysRevLett.114.016401}.
After atomic relaxation, our calculations find that 
the (SrTaO$_3$)$_1$/(SrIrO$_3$)$_1$ superlattice is 
stabilized in the 
aforementioned $P4/mbm$ $\cal{SG}$ crystal structure, 
which is characterized by an 
out-of-phase OO rotation. Ta$^{4+}$ has a nominal $d^1$ 
occupancy and Ir$^{4+}$ has a nominal $d^5$ occupancy.
Panel \textbf{a} of Fig.~\ref{fig2} shows the density of states of the
superlattice. The shades and the solid curves correspond to the densities 
of states (DOS) calculated using DFT and DFT+SOC, respectively.
The gray, red and green shades/curves are total, Ta-$d$ projected 
and Ir-$d$ projected densities of states, respectively. 
Due to electronegativity difference, 
Ta-$d$ states have higher energy than Ir-$d$ states, but they have 
overlap around the Fermi level. Panel \textbf{b} shows the DFT  
band structure of the (SrTaO$_3$)$_1$/(SrIrO$_3$)$_1$ superlattice 
close to the Fermi level without SOC. 
The red curves are Ta-$d$ projected bands and the green curves are 
Ir-$d$ projected bands. Clearly there are band inversions between 
Ta-$d$ bands and Ir-$d$ bands.
Without SOC the highest valence band and the lowest conduction band 
form a node 
surface in the Brillouin zone.
Panel \textbf{c} shows the DFT+SOC band structure of the superlattice 
close to the Fermi level. With SOC, a full gap is opened 
between the highest valence band and the lowest conduction band
in the entire Brillouin zone. Because the crystal structure of the
(SrTaO$_3$)$_1$/(SrIrO$_3$)$_1$ superlattice has inversion symmetry, 
we can use the parity rule to easily calculate the $Z_2$ topological 
index~\cite{10.1103/PhysRevLett.98.106803}. Panel \textbf{d} 
shows $\frac{1}{8}$ of the Brillouin zone with the eight time-reversal 
invariant momentum (TRIM) points explicitly labelled. 
The parity of each TRIM point is shown in the figure. 
We find that all the TRIM points have 
even parity except that Z point has odd parity.
Therefore the $Z_2$ index of the superlattice 
is (1;001), which is a strong topological insulator. Panel \textbf{d} 
also shows the (100)-projected and (001)-projected surface 
Brillouin zones. Using the Green function method, we calculate the 
(100) surface bands and (001) surface bands in panels \textbf{e} 
and \textbf{f}, respectively. In both cases, the topologically 
protected surface bands traverse the band gap. For the 
(100) surface, the crossing of the surface bands at 
$\overline{\textrm{Z}}$ point 
occurs in the gap and close to the Fermi level (see panel \textbf{e}). 
However, for the (001) surface, the crossing of the surface bands at 
$\overline{\Gamma}$ point is well below the Fermi level. 

\begin{figure}
    \centering
    \includegraphics[width=0.8\textwidth]{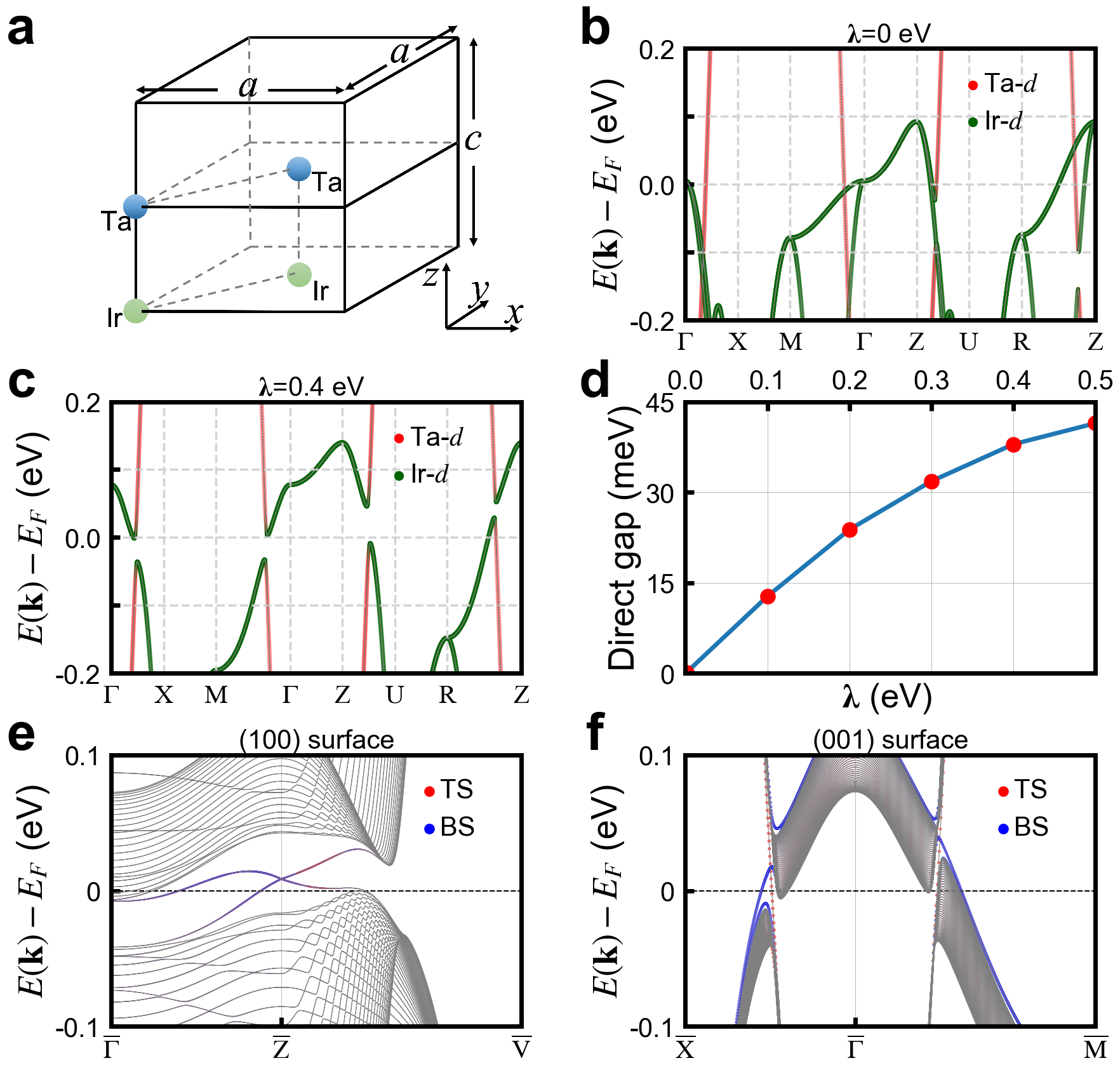}
    \caption{\textbf{Electronic and topological properties of the
        (SrTaO$_3$)$_1$/(SrIrO$_3$)$_1$ superlattice from model
        calculations}. \textbf{a}: The lattice sites for Ta and Ir atoms
      on which a simple tight-binding Hamiltonian model
      $H_{\lambda}(\textbf{k})$ is built.  The blue and green balls
      represent Ta and Ir atoms, respectively.  \textbf{b}: Near Fermi
      level band structure of the tight-binding model
      $H_{\lambda}(\textbf{k})$ with $\lambda = 0$ in the bulk
      Brillouin zone. The red and green symbols highlight the atomic
      projection onto Ta-\textit{d} and Ir-\textit{d} orbitals,
      respectively. The Fermi level is determined by the total
      occupancy of $N=12$ and is shifted to the zero energy. Gap is
      not opened around the Fermi level.  \textbf{c}: Same as panel
      \textbf{b} except that $\lambda$ = 0.4 eV in
      $H_{\lambda}(\textbf{k})$. A gap is opened along the
      high-symmetry \textbf{k}-path. \textbf{d}: The smallest direct
      band gap of $H_{\lambda}(\textbf{k})$ as a function of
      $\lambda$. The number of valence bands is equal to the total
      occupancy of $N=12$. \textbf{e}: Band structure of (100)
      semi-infinite slab of the tight-binding model
      $H_{\lambda}(\textbf{k})$ with $\lambda =$ 0.4 eV, calculated by
      using the exact diagonalization method. The red and blue symbols
      are the projections onto the top surface (TS) and the bottom
      surface (BS).  \textbf{f}: Same as panel \textbf{e} except for
      (001) semi-infinite slab of the tight-binding model
      $H_{\lambda}(\textbf{k})$ with $\lambda =$ 0.4 eV.}
    \label{fig3}
\end{figure}

To understand why non-trivial topology emerges in the band 
structure of the (SrTaO$_3$)$_1$/ (SrIrO$_3$)$_1$ superlattice, 
we construct a tight-binding model with the aid of maximally 
localized Wannier functions. We
focus on Ta and Ir atoms, because
there are only Ta-$d$ and Ir-$d$ bands around the Fermi 
level. Ta atoms and Ir atoms are on a respective 
two-dimensional square lattice, which alternates with each other 
along the $z$ axis (see panel \textbf{a} of Fig.~\ref{fig3}).  
Using the current $(xyz)$ coordinate system in panel \textbf{a}, 
we consider Ta-$d_{xz}$, $d_{yz}$, $d_{x^2-y^2}$ orbitals and 
all five Ir-$d$ orbitals (altogether $(3+5)\times 2 = 16$ atomic orbitals).  
The tight-binding Hamiltonian $H_{\lambda}(\textbf{k})$ is:
\begin{equation}
    \label{eq1} H_{\lambda}(\textbf{k}) = H_0(\textbf{k})\otimes I_2 + 
    H_{\textrm{SOC}} 
\end{equation}
where $H_0(\textbf{k})$ is a $16 \times 16$ matrix which 
fits the DFT band structure without SOC, $I_2$ is a 
$2\times 2$ identity matrix which is in the spin basis and 
$H_{\textrm{SOC}}$ is a $32 \times 32$ matrix that 
describes the atomic spin-orbit interaction. $H_0(\textbf{k})$ 
can be obtained via maximally localized Wannier function~\cite{10.1103/RevModPhys.84.1419}
(see Supplementary Notes 4). 
$H_{\textrm{SOC}} = \lambda \textbf{L}\cdot \textbf{S}$ 
where $\lambda$ is the atomic spin-orbit coupling constant (the 
explicit form of $H_{\textrm{SOC}}$ is shown in Supplementary Notes 6). 
For simplicity, we 
assume a common spin-orbit coupling constant $\lambda$, since Ta and Ir 
have close atomic numbers (73 versus 77). The advantage of 
this simple tight-binding model is that SOC is not absorbed in the 
basis and thus we can study the SOC effect explicitly. Panel \textbf{b} 
shows the band structure of $H_{\lambda}(\textbf{k})$ with $\lambda = 0$ (i.e. no 
SOC included). We find that with the aid of maximal localized Wannier 
function, $H_0(\textbf{k})$ almost exactly reproduces the DFT band structure 
without SOC. The band inversion between Ta-$d$ (red symbols) 
and Ir-$d$ (green symbols) 
states is clearly seen. Panel \textbf{c} shows the band structure of 
the full Hamiltonian 
$H_{\lambda}(\textbf{k})$ with $\lambda = 0.4$ eV. 
We find that with SOC included, a full gap is opened between the 
highest valence band and the lowest conduction band around the Fermi level, 
which is very similar to the DFT+SOC result (see Fig.~\ref{fig2}\textbf{c}).
Once the gap is opened with a positive $\lambda$, we can 
calculate the parity of all the valence bands at TRIM points. 
It turns out that they are identical to those from the DFT+SOC 
calculations, i.e. a strong topological 
insulator. In addition, we calculate the minimum direct 
gap $\Delta$ in the Brillouin zone as a function of $\lambda$. As panel \textbf{d} 
shows, $\Delta$ monotonically increases with $\lambda$ (up to 0.5 eV).
Based on the model Hamiltonian $H_{\lambda}(\textbf{k})$, we can 
calculate the surface 
bands using the Green function method. Panels \textbf{e} and \textbf{f} 
show the (100) and (001) surface bands, respectively. The blue and red 
dots highlight the projections onto the top surface (TS) and the bottom 
surface (BS). The surface bands from the model calculations are also 
very similar to the first-principles calculations. After verifying that 
the simple tight-binding model is adequate to describe the band 
topology of the (SrTaO$_3$)$_1$/(SrIrO$_3$)$_1$ superlattice, we 
now elucidate why non-trivial band topology emerges in this system. It 
depends on several factors.
We first count the occupancy: Ta has a $d^1$ occupancy and Ir has a 
$d^5$ occupancy. Therefore in the model the total occupancy is 
$(1+5)\times2 = 12$. Next we note that all the bands are four-fold 
degenerate at the TRIM points X, Y, M, U, V, R. 
This is due to the combination of time-reversal and symmetry operations 
in $P4/mbm$ $\cal{SG}$ (see the symmetry analysis of $P4/mbm$ $\cal{SG}$ in 
Supplementary Notes 1). 
Considering that the total occupancy 
is 12, the parity of X, Y, M, U, V, R must be even. 
Therefore the non-trivial band topology arises from the fact that 
$\Gamma$ has even parity and Z has odd parity. However, band inversion 
occurs at both $\Gamma$ and Z points. We explain that while both 
TRIM points have band inversion, the even parity of $\Gamma$ and the 
odd parity of Z are closely 
related to the special coordinates of Ir and Ta in the superlattice.
As panel \textbf{a} shows, Ir atoms are at $(0, 0, 0)$ and 
$(\frac{1}{2}, \frac{1}{2},0)$ in fractional coordinate system, while 
Ta atoms are at $(0,0,\frac{1}{2})$ and 
$(\frac{1}{2},\frac{1}{2},\frac{1}{2})$. The inversion center is at 
$(0,0,0)$, with respect to which inversion operation leads to 
even number shifts of lattice vectors for Ir atoms, but odd number shifts 
of lattice vectors for Ta atoms. Consequently, the phase shift induced 
to the Bloch wavefunction is always 0 for Ir-$d$ states at $\Gamma$ and 
Z, but for Ta-$d$ states the phase shift is 0 at $\Gamma$ and $\pi$ at Z. 
We have band inversion between Ta-$d$ and Ir-$d$ bands 
at both $\Gamma$ and Z points. At $\Gamma$, band inversion does not 
change the parity, which is always even. However, when Ta-$d$ 
band and Ir-$d$ band invert at Z, the parity switches and as a 
result, non-trivial band topology emerges. In summary,
the origin of non-trivial topology precisely lies in the special
atomic positions of $B$ and $B'$ in the ($AM$O$_3$)$_1$/($AM'$O$_3$)$_1$
superlattice, where the wave-functions of $M$ and $M'$ atoms
gain different phase shift under inversion operation.

We comment that while we use the 
inversion symmetry and the parity rule to calculate the $Z_2$ topological 
index, we can explicitly show that the strong topological insulating state 
in the (SrTaO$_3$)$_1$/(SrIrO$_3$)$_1$ superlattice
does not depend on the inversion symmetry and/or $C_4$ rotation symmetry 
associated with the $P4/mbm$ $\cal{SG}$. We can artificially move the 
Ta or Ir atoms in the superlattice so that the inversion symmetry or 
$C_4$ rotation symmetry is explicitly broken. As long as a gap is opened between 
the highest valence band and the lowest conduction band throughout the 
Brillouin zone, we can use the Wilson loop method~\cite{10.1103/PhysRevB.84.075119} to calculate the $Z_2$ topological index, which still turns out to be $(1;001)$.

\subsection*{Coexisting TI and TDS states in the (SrMoO$_3$)$_1$/(SrIrO$_3$)$_1$ superlattice}

\begin{figure}
    \centering
    \includegraphics[width=0.8\textwidth]{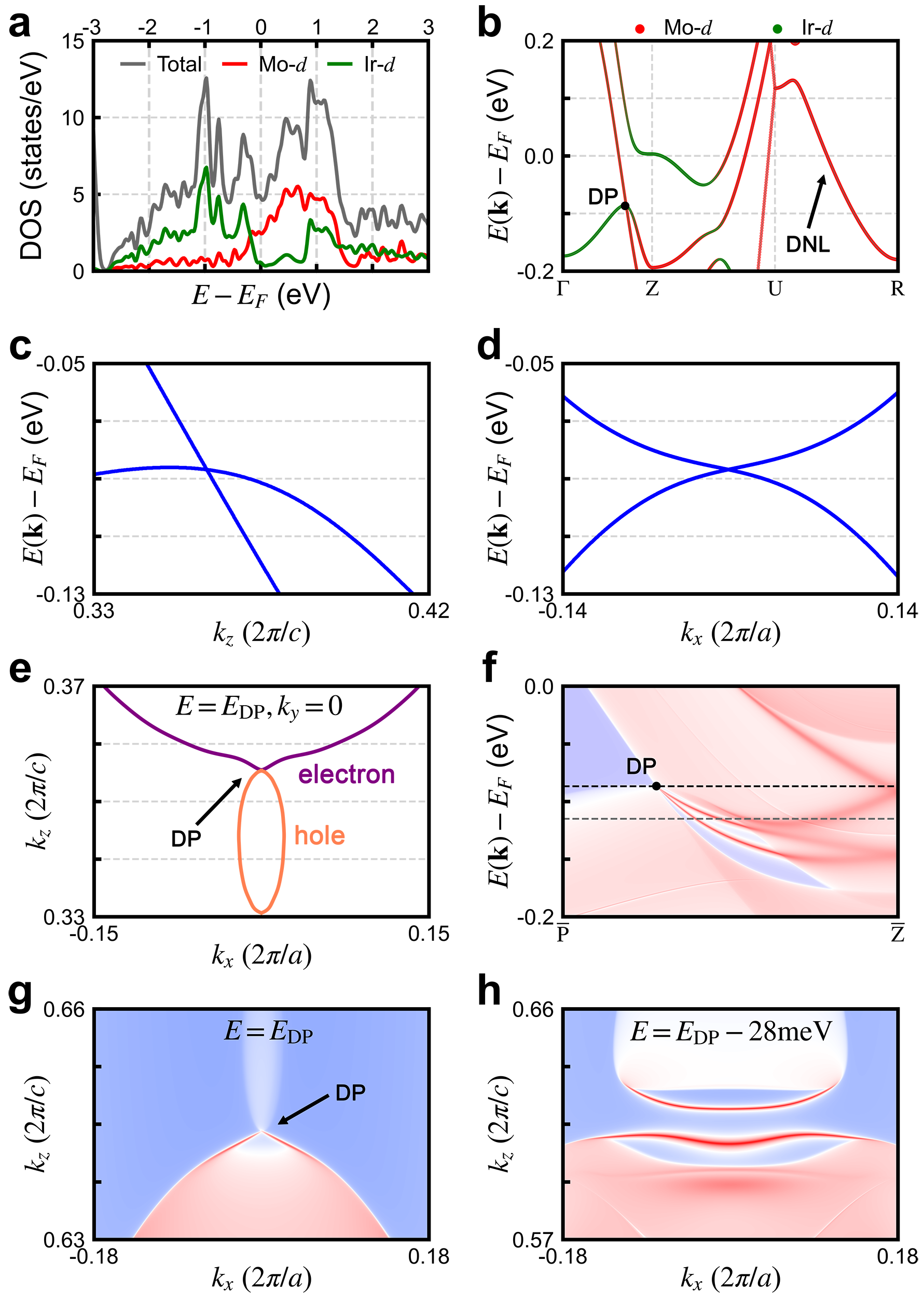}
    \caption{\textbf{Electronic and topological properties of the
        (SrMoO$_3$)$_1$/(SrIrO$_3$)$_1$ superlattice from
        first-principles calculations}. \textbf{a}: Density of states
      (DOS) of the (SrMoO$_3$)$_1$/(SrIrO$_3$)$_1$ superlattice,
      calculated by DFT+SOC method.  The gray, red and green curves
      are total, Mo-$d$ projected and Ir-$d$ projected DOS.
      \textbf{b}: Near Fermi level DFT+SOC band structure of the
      superlattice.  The red and green symbols are Mo-$d$ and Ir-$d$
      projected orbitals, respectively.
      The black dot highlights a Dirac point (DP)
      along the $\Gamma \to$ Z path. The black arrow highlights a
      Dirac node-line}
    \label{fig4}
\end{figure}

In this section, we study a $d^2+d^5$ system: the
(SrMoO$_3$)$_1$/(SrIrO$_3$)$_1$ superlattice and demonstrate that
topological Dirac semi-metal state (TDS) and topological insulator
state (TI) coexist in the superlattice. Similar to bulk SrTaO$_3$,
SrMoO$_3$ is non-magnetic and crystallizes in a cubic structure
($\cal{SG}$ $Pm\bar{3}m$,
No. 221)~\cite{10.1016/j.jssc.2009.11.005}. After atomic relaxation,
our calculations find that the (SrMoO$_3$)$_1$/(SrIrO$_3$)$_1$
superlattice also crystallizes in the $P4/mbm$ structure ($\cal{SG}$
No. 127), similar to the (SrTaO$_3$)$_1$/(SrIrO$_3$)$_1$ superlattice.
We first consider KTaO$_3$ substrate, which has a very small lattice
mismatch with the (SrMoO$_3$)$_1$/(SrIrO$_3$)$_1$ superlattice (the
strain effect will be discussed later in this section).  Panel
\textbf{a} of Fig.~\ref{fig4} shows the DFT+SOC density of states of
the (SrMoO$_3$)$_1$/(SrIrO$_3$)$_1$ superlattice. Because the
electronegativity difference between Mo-$d$ and Ir-$d$ states is
smaller than that between Ta-$d$ and Ir-$d$
states~\cite{10.1088/1361-648X/aa6efe}, we can see that Mo-$d$ and
Ir-$d$ states have more overlaps around the Fermi level. Panel
\textbf{b} of Fig.~\ref{fig4} shows the DFT+SOC band structure around
the Fermi level. We find that the highest valence band and the lowest
conduction band cross along $\Gamma$-Z path. The crossing point is
close to the Fermi level (87 meV below the Fermi level). This DP
emerges in pairs and only along $\Gamma$-Z path (0,0,$k_z$) and
(0,0,$-k_z$), because of the combination of time-reversal
$\mathcal{T}$, inversion $\mathcal{I}$ and $C_4$ rotation symmetry
(associated with $P4/mbm$ $\cal{SG}$)~\cite{10.1038/ncomms5898}.  The
highest valence band is mainly composed of Ir-$d$ orbital and the
lowest conduction band is mainly composed of Mo-$d$ orbital. The two
bands have different eigenvalues of the $C_4$ rotation operation and
thus they cannot hybridize with each other at the crossing point.
Therefore the DP is stabilized even in the presence of SOC.  However,
while the highest valence band and the lowest conduction do not mix
along $\Gamma$-Z, it is not necessary that the two bands invert. The
band inversion in the (SrMoO$_3$)$_1$/(SrIrO$_3$)$_1$ superlattice is
a consequence of proper electronegativity difference between Mo and
Ir, which can be controlled by external parameters such as epitaxial
strain (see discussion below).  In addition to the pair of DP, we also
find Fermi level crosses a DNL along U-R (see panel \textbf{b}). This
DNL is protected by the non-symmorphic symmetry associated with
$P4/mbm$ $\cal{SG}$ (No. 127) together with inversion and
time-reversal symmetries.  Oxygen octahedral rotations in the crystal
structure leads to band folding, which leads to a four-fold degeneracy
of all Bloch states along X-M-Y and U-R-V (see symmetry analysis of
$P4/mbm$ $\cal{SG}$ in Supplementary Notes 1). The Fermi level must
cross one of those DNL because of the total occupancy. Mo has a $d^2$
occupancy and Ir has a $d^5$ occupancy. Hence the total occupancy is
$(2+5)\times 2 = 14$, which is $3\times 4 + 2$ and thus one band must
be half-filled. We note that such DNL also exists in the
(SrTaO$_3$)$_1$/(SrIrO$_3$)$_1$ superlattice, since the two
superlattices have the same symmetries. However, the
(SrTaO$_3$)$_1$/(SrIrO$_3$)$_1$ superlattice has a total occupancy of
12, which can be divided by 4. Thus the Fermi level does not cross the
DNL of the (SrTaO$_3$)$_1$/(SrIrO$_3$)$_1$ superlattice.

Next we analyze the DP in more details. Panels \textbf{c} and 
\textbf{d} of Fig.~\ref{fig4} show the band structure around the DP along $k_z$ and 
$k_x$ axes. We find that in the vicinity of the DP, the Fermi velocities of the two 
bands have the same signs along $k_z$ axis, but have opposite signs along 
$k_x$ axis. This suggests that the DP in the (SrMoO$_3$)$_1$/(SrIrO$_3$)$_1$ 
superlattice is of type-II~\cite{10.1103/PhysRevLett.119.026404}. To confirm that, we plot a constant energy 
contour in the $k_y = 0$ plane $(k_x, k_z)$ in the Brillouin zone 
with the energy set to be that of the DP (see panel \textbf{e}). 
We find that the DP is a 
touching point between an electron pocket and a hole pocket, which is the evidence 
of type-II DP. Furthermore we calculate the topological properties of the DP. 
We find that the 2D topological invariant $\nu_{\textrm{2D}}$ = 0 in both $k_z =0$ 
and $k_z = \pi$ planes. However, the mirror Chern number $n_{\mathcal{M}}$ 
of the $k_z = 0$ plane is equal to 2. These topological properties are distinct 
from existing TDS Na$_3$Bi and Cd$_3$As$_2$ ($\nu_{\textrm{2D}}$ = 1 and 
$n_{\mathcal{M}} = 1$)~\cite{10.1103/PhysRevB.85.195320,10.1038/ncomms4786,10.1103/PhysRevLett.113.027603,10.1103/PhysRevB.88.125427} but 
identical to VAl$_3$~\cite{10.1103/PhysRevLett.119.026404}.
The non-trivial $k_z=0$ mirror Chern number $n_{\mathcal{M}}$ manifests 
itself as the number of Fermi arc associated with the DP. 
Panel \textbf{f} shows the surface bands along a high-symmetry \textbf{k}-path in the (010) 
surface Brillouin zone. Since the DP is along $\Gamma$ to Z, when it is projected 
onto the (010) surface Brillouin zone, the projection is along $\overline{\Gamma}$ 
to $\overline{\textrm{Z}}$. Panel \textbf{f} shows the surface bands of the 
(SrMoO$_3$)$_1$/(SrIrO$_3$)$_1$ superlattice from $\overline{\textrm{P}}$ (0, 0.3) 
to $\overline{\textrm{Z}}$. 
We can clearly see that two Fermi arcs emerge from the projection 
of the DP at $(0, 0.36)$. They will terminate at the projection of the 
other DP at $(0, -0.36)$. Panels \textbf{g} and \textbf{h} show the constant 
energy contour of the surface bands in the 
(010) surface Brillouin zone $(k_x,k_z)$. The constant energy value of panel 
\textbf{g} is that of 
the DP $E=E_{\textrm{DP}}$ (see the black dashed line in panel \textbf{f}). The 
projection of the DP is highlighted by a black arrow.
The constant energy value of panel \textbf{h} is $E = E_{\textrm{DP}} - 28$ meV (see the gray 
dashed line in panel \textbf{f}). At this energy, the two Fermi arcs are clearly visible
in the (010) surface Brillouin zone. 

As we mentioned above, the symmetries (time reversal + inversion + $C_4$ rotation) allow 
DP to occur, but the actual emergence of DP depends on band inversion. We find 
that epitaxial strain can control the band inversion and thus the number of DP pairs and 
their positions in the Brillouin zone. Panels \textbf{a}, \textbf{b} and \textbf{c} of 
Fig.~\ref{fig5} show the DFT+SOC band structure of the (SrMoO$_3$)$_1$/(SrIrO$_3$)$_1$ 
superlattice under 1\%, 2\% and 3\% compressive strain, respectively. We choose the 
\textbf{k}-path along $\Gamma$ to Z because DP only appears on that \textbf{k}-path. 
We find that strain can change the relative energy 
position of the lowest conduction band (blue) and the highest valence band (red)
and thus their crossing points close to the Fermi level. When the superlattice 
is under 1\% compressive strain, there is one pair of DP; when it is under 2\% 
compressive strain, there are two pairs of DP; when it is under 3\% compressive 
strain, there are no band crossings and thus no DP. We find that the mirror Chern number 
of the DP is 2. Panel \textbf{d} shows the 
phase diagram of the (SrMoO$_3$)$_1$/(SrIrO$_3$)$_1$ superlattice as a function of 
epitaxial strain. From 2\% tensile strain to 1.7\% compressive strain, the superlattice 
has one pair of DP. In a narrow compressive strain range (from 1.7\% to 2.7\%), 
the superlattice has two pairs of DP. When the compressive strain exceeds 2.7\%, 
the superlattice does not have DP. KTaO$_3$ has negligible lattice mismatch to 
the (SrMoO$_3$)$_1$/(SrIrO$_3$)$_1$ superlattice and therefore the superlattice 
on a KTaO$_3$ substrate has one pair of DP. The more widely used substrate 
SrTiO$_3$ imposes a 2.3\% compressive strain on the (SrMoO$_3$)$_1$/(SrIrO$_3$)$_1$ 
superlattice and thus the superlattice on a SrTiO$_3$ substrate has two pairs of DP. 
However, in all the above cases, DNL still exists and the Fermi level must cross a 
DNL, which is guaranteed by the non-symmorphic symmetry and the total occupancy. 
Similar to the previous (SrTaO$_3$)$_1$/(SrIrO$_3$)$_1$ superlattice, we can also build a 
tight-binding Hamiltonian $H(\textbf{k}) = H_0(\textbf{k})\otimes I_2 + H_{\textrm{SOC}}$ 
to study the SOC effects on the DP (see Supplementary Notes 7). 
We find that theoretically, 
tuning the strength of SOC can also control the band inversion and thus the number of DP pairs. 
For a range of SOC strength (reasonable for Mo and Ir), the tight-binding model finds 
one pair of DP along $\Gamma$ to Z.

Finally, we demonstrate the peculiar topological properties of the
(SrMoO$_3$)$_1$/(SrIrO$_3$)$_1$ superlattice: the coexistence of TDS
and TI.  Panel \textbf{a} of Fig.~\ref{fig6} shows the
near-Fermi-level band structure of the (SrMoO$_3$)$_1$/(SrIrO$_3$)$_1$
superlattice.  The highest valence band and the lowest conduction band
have a linear crossing along $\Gamma$ to Z with a non-trivial mirror
Chern number. This leads to the aforementioned TDS state, which is
highlighted by the light red shade in
Fig.~\ref{fig6}\textbf{a}.  However, the highest valence band
and the second highest valence band, as well as the lowest conduction
band and the second lowest conduction band do not have crossing in the
entire Brillouin zone and the gap between them (highlighted by the
light yellow shades) turns out to be topologically nontrivial with
$Z_2$ index (1;001).  The emergence of those TI states has the same
origin as the (SrTaO$_3$)$_1$/(SrIrO$_3$)$_1$ superlattice.  Panel
\textbf{b} shows the (010) projected surface bands of the
(SrMoO$_3$)$_1$/(SrIrO$_3$)$_1$ superlattice.  The high-symmetry
\textbf{k}-path is taken from $\overline{\textrm{P}}$ (0, 0.3) to
$\overline{\textrm{Z}}$ to $\overline{\textrm{P}}$ in the 2D surface
Brillouin zone.  The red highlights the projection of the surface
bands. We find that the TDS surface Dirac cone is sandwiched between
two TI surface Dirac cone in the energy-momentum space. In particular,
the upper TI surface Dirac cone crosses the Fermi level. With hole
doping, the Fermi level may be shifted so as to cross the TDS surface
Dirac cone and the lower TI surface Dirac cone. Such a TI-TDS-TI
topological phase transition may be achieved via chemical
doping~\cite{10.1038/srep14564} or electric-field gating~\cite{10.1038/s41427-018-0062-1}.
The coexistence of a TI surface Dirac cone and a TDS
  surface Dirac cone near the Fermi level was also found in iron-based
  superconductors such as FeSe~\cite{10.1038/s41567-018-0280-z}. In those materials,
  the band inversion is between Fe-$d_{xz/yz}$ orbitals and
  ligand-$p_z$ orbitals and the non-trivial topology arises because of
  the dissimilar orbital parity (a different mechanism from the one in
  this work). Furthermore in our case of the
  (SrMoO$_3$)$_1$/(SrIrO$_3$)$_1$ superlattice, the topological
  surface bands are richer in that there are two TI surface Dirac
  cones, which sandwich a TDS surface Dirac cone.

\section*{Discussion}

Before we conclude, we make a few comments. First,
  iridates such as Sr$_2$IrO$_4$ and (SrTiO$_3$)$_1$/(SrIrO$_3$)$_1$
  superlattice exhibits canted antiferromagnetism (AFM) at low
  temperatures~\cite{10.1103/PhysRevB.49.9198,10.1103/PhysRevLett.108.247212,10.1103/PhysRevLett.101.076402,10.1103/PhysRevLett.114.247209}. We
  investigate whether canted AFM may also emerge in the
  (SrTaO$_3$)$_1$/ (SrIrO$_3$)$_1$ and (SrMoO$_3$)$_1$/(SrIrO$_3$)$_1$
  superlattices. From DFT+$U$+SOC calculations, we find that up to
  $U_{\rm{Ir}}$ of 4 eV, no canted AFM is stabilized in either
  superlattice and thus their topological properties remain robust
  (see Supplementary Notes 5).  In literature,
  considering that Ir is a third-row transition metal element with
  more extended $5d$ orbitals, its effective Hubbard $U$ ranges
  between 1 and 3 eV~\cite{10.1103/PhysRevB.99.245145,10.1073/pnas.1907043116,10.1103/PhysRevB.92.054428}. Second, it is
well known that DFT calculations with semi-local exchange-correlation
functionals underestimate the oxygen-$p$ and metal-$d$ separation
($p$-$d$ separation) in complex oxides. However, in our study, we
focus on $d$-$d$ band inversion. While the $p$-$d$ separation is
underestimated in each Sr$M$O$_3$, the $d$-$d$ separation is less
affected by the semi-local exchange-correlation functionals, as
demonstrated in the previous study~\cite{10.1103/PhysRevB.90.245138}.
Third, both in the (SrTaO$_3$)$_1$/(SrIrO$_3$)$_1$ and
  (SrMoO$_3$)$_1$/(SrIrO$_3$)$_1$ superlattices, their (100) surface
  exhibits topologically non-trivial surface states. While the (001)
  surface is the most natural surface for a (001)-grown oxide
  superlattice, a (100) or (010) surface can also be obtained by
  either cleaving the sample or mechanically polishing the
  sample~\cite{10.1126/science.1165044,10.1016/j.scib.2018.10.001}. Such preparation of (100) or
  (010) surfaces is routinely made in transmission electron microscopy
  (TEM) and cross-sectional scanning tunneling microscopy (STM)
  measurements~\cite{10.1021/acsami.9b14641,10.1126/sciadv.abj0481,10.1021/acsnano.6b08260,10.1021/acsami.7b00150}. Finally, we
  comment on how one may separate the topological surface bands from
  the bulk spectrum. We notice that in both superlattices
  ((SrTaO$_3$)$_1$/(SrIrO$_3$)$_1$ and (SrMoO$_3$)$_1$/(SrIrO$_3$)$_1$),
  the topological surface states
  overlap with the bulk spectrum in the energy window. However, as
  Fig.~\ref{fig2}\textbf{e} and Fig~\ref{fig4}\textbf{f} show, in the
  energy window that is close to the Fermi level, the topological
  surface bands and the bulk spectrum are separated in the momentum
  space. This enables direct spectroscopic observation (such as ARPES)~\cite{10.1126/science.1173034,10.1038/nmat3990}. In addition, in
  actual experiments, disorders are inevitable, which may suppress the
  bulk conduction via the Anderson localization~\cite{10.1103/PhysRev.109.1492}. By contrast, the conduction from the
  topological surface bands can evade the Anderson localization (as
  long as the defect is non-magnetic)~\cite{10.1103/RevModPhys.82.3045}. This scenario likely occurs to
  the (SrTaO$_3$)$_1$/(SrIrO$_3$)$_1$ superlattice, whose bulk Fermi
  surface consists of small electron and hole pockets. Finally, while
  the bulk topological properties protect the \textit{presence} of the
  surface states~\cite{10.1103/RevModPhys.82.3045}, the
  details of the surface states can also be tuned in
  experiment. Adsorption and removal of non-magnetic atoms on the
  surface as well as different surface terminations can modify the curvature of
  the topological surface bands, which provides another means to
  separate the surface states from the bulk spectrum~\cite{10.1021/nn200556h,10.1038/ncomms1131,10.1016/j.physleta.2011.12.022,https://arxiv.org/abs/1012.2927,https://arxiv.org/abs/1105.4794}.

In summary, we combine first-principles calculations and tight-binding 
models to show that between the highest valence band and the lowest conduction 
band, the (SrTaO$_3$)$_1$/ (SrIrO$_3$)$_1$ superlattice 
exhibits a strong topological insulating state (TI) and the 
(SrMoO$_3$)$_1$/ (SrIrO$_3$)$_1$ superlattice is a topological Dirac semi-metal (TDS)
that is characterized by a pair of type-II DP with a mirror Chern number of 2, as well 
as DNL protected by nonsymmorphic space group. 
The (SrMoO$_3$)$_1$/(SrIrO$_3$)$_1$ superlattice also 
exhibits a strong topological insulating state between the highest 
valence band and the second highest valence band, as well as the lowest 
conduction band and the second lowest conduction band. The coexistence of multiple 
topological states in the (SrMoO$_3$)$_1$/(SrIrO$_3$)$_1$ superlattice leads to 
a characteristic 'sandwich' structure of the surface bands:
one TDS surface Dirac cone lies between two TI surface Dirac cones in the 
energy-momentum space. For both TI and TDS states, the 
non-trivial band topology arises from band inversion between 
$d$ states of two different transition metal atoms, combined with 
proper $d$ occupancy and time-reversal symmetry. In the TDS state, 
crystal symmetry ($C_4$ rotation symmetry and nonsymmorphic $P4/mbm$ space 
group) also plays a crucial role. The design principle is also applicable to 
first-row transition metal atoms whose $d$ orbitals are more strongly 
correlated and may break time-reversal symmetry with long-range magnetic 
order. In a (001) ($AM$O$_3$)/($AM'$O$_3$) oxide superlattice with 
stronger correlation effects on transition metal $d$ orbitals, other 
topological states such as quantum anomalous Hall state and Weyl semi-metal state 
may emerge. This will be in our future research.

\section*{Methods}

We perform density functional theory (DFT)~\cite{10.1103/PhysRev.136.B864,10.1103/PhysRev.140.A1133} 
calculations, as implemented in 
Vienna Ab Initio Simulation Package (VASP)~\cite{10.1103/PhysRevB.49.14251,10.1103/PhysRevB.54.11169}. 
We use the generalized 
gradient approximation with the Perdew–
Burke–Ernzerhof parameterization 
revised for solids (PBEsol)~\cite{10.1103/PhysRevLett.100.136406} 
as the exchange-correlation functional. We use an energy cutoff of 600 eV and a 
$10\times10\times8$ Monkhorst-Pack \textbf{k}-mesh to sample the 
Brillouin zone~\cite{10.1103/PhysRevB.13.5188} of 
the superlattice. Spin-orbit coupling (SOC) is self-consistently 
included in all the calculations 
unless otherwise specified. The convergence threshold for the self-consistent 
calculation is $10^{-6}$ eV. Atomic relaxation is converged when each force component 
is smaller than 0.01 eV $\mathrm{\AA}$$^{-1}$ and pressure on the simulation cell is less 
than 0.5 kbar. For bi-axial strain calculations, we fix the two in-plane lattice 
constants and allow the out-of-plane lattice constant (along the stacking direction) 
to fully relax. The strain is defined as $\xi = (a-a_0)/a_0 \times 100\%$ where 
$a_0$ is the DFT optimized pseudo-tetragonal lattice constant and $a$ is the 
theoretical lattice constant of the substrate.

We employ the Wannier90 packages~\cite{10.1016/j.cpc.2007.11.016} to construct the maximally 
localized Wannier functions (MLWF)~\cite{10.1103/RevModPhys.84.1419}. We use two sets 
of Wannier functions. 
The first set of Wannier functions is to 
fit the band structure with SOC. This set of Wannier functions, combined with 
Green function method, is used to calculate the surface bands, as implemented in  
WannierTools~\cite{10.1016/j.cpc.2017.09.033}. 
The second set of Wannier functions is to 
fit the band structure without SOC. This set of Wannier 
function is used to construct the tight-binding model in which 
atomic SOC interaction is explicitly shown in an 
analytical form and studied.

\section*{Data availability }
The data that support the findings of this study are available from the corresponding author upon reasonable request.

\section*{Code availabilty}
The electronic structure calculations were performed using the
proprietary code VASP~\cite{10.1103/PhysRevB.54.11169}, the open-source codes Wannier90~\cite{10.1016/j.cpc.2007.11.016} and WannierTools~\cite{10.1016/j.cpc.2017.09.033}. Both
Wannier90 and WannierTools are freely distributed on
academic use under the Massachusetts Institute of Technology (MIT)
License.

\section*{Acknowledgement}
We are grateful to Hongming Weng, Haizhou Lu, Zhijun Wang, Qihang Liu,
Jiachen Gao and Yuhao Gu for useful discussions and in particular to
Andrew Wray for his critical reading of our manuscript. H.C. is
supported by the National Natural Science Foundation of China under
project number 11774236, the Ministry of Science and Technology of
China under project number SQ2020YFE010418, and Open Grant of State
Key Laboratory of Low Dimensional Quantum Physics at Tsinghua
University.  G.L. is supported by NSF of China (Grant No. 11874263),
the National Key R$\&$D Program of China (2017YFE0131300), and
Shanghai Technology Innovation Action Plan 2020-Integrated Circuit
Technology Support Program (Project No. 20DZ1100605).

\section*{Author contributions}
H.C. conceived and supervised the project. Z.L. and H.L. contributed equally to this work. Z.L. and H.C. performed the first-principles calculations. G.L. and H.L. performed the model calculations. G.L. refined the focus of the manuscript with H.C.. J.M. and X.W. participated in the data analysis. H.C., G.L., Z.L. and H.L. wrote the manuscript.

\section*{Competing interests}
The authors declare no competing financial or non-financial interests.

\clearpage
\newpage




\begin{figure}
 \justifying\noindent$\displaystyle${(DNL) along the U $\to$ R
   path. \textbf{c}: Band dispersion in the vicinity of the DP along
   $k_z$ direction. \textbf{d}: Band dispersion in the vicinity of
   the DP along $k_x$ direction. \textbf{e}: Constant energy at the
   $k_y=0$ $(k_x, k_z)$ plane in the bulk Brillouin zone.  The
   constant energy is chosen as the energy of the Dirac point. The
   purple (orange) curve highlights that it encloses an electron
   (hole) pocket. The Dirac point emerges where the electron and hole
   pockets touch each other, which is highlighted by the black
   arrow. \textbf{f}: Band structure of (010) semi-infinite slab of
   the (SrMoO$_3$)$_1$/(SrIrO$_3$)$_1$ superlattice, calculated by
   using the Green-function method.  The \textbf{k}-path in the
   surface Brillouin zone is taken to be
   $\mathrm{\overline{P}}\to\mathrm{\overline{Z}}$, where
   $\mathrm{\overline{P}}$ is $(0.0, 0.3)$ and
   $\mathrm{\overline{Z}}$ is $(0.0, 0.5)$. The black dash line
   highlights the Dirac point energy. The gray dashed line highlights
   the energy that is 28 meV below the Dirac point
   energy. \textbf{g}: A constant energy contour of band structure of
   (010) semi-infinite slab of the (SrMoO$_3$)$_1$/(SrIrO$_3$)$_1$
   superlattice in the (010) surface Brillouin zone. The constant
   energy is the energy of DP $E= E_{\textrm{DP}}$, shown by the
   black line in panel \textbf{f}.  \textbf{h}: The same as panel
   \textbf{g}, except that the constant energy is $E =
   E_{\textrm{DP}} - 28$ meV, shown by the gray line in panel
   \textbf{f}.}
\end{figure}

\newpage
\clearpage

\begin{figure}
    \centering
    \includegraphics[width=0.8\textwidth]{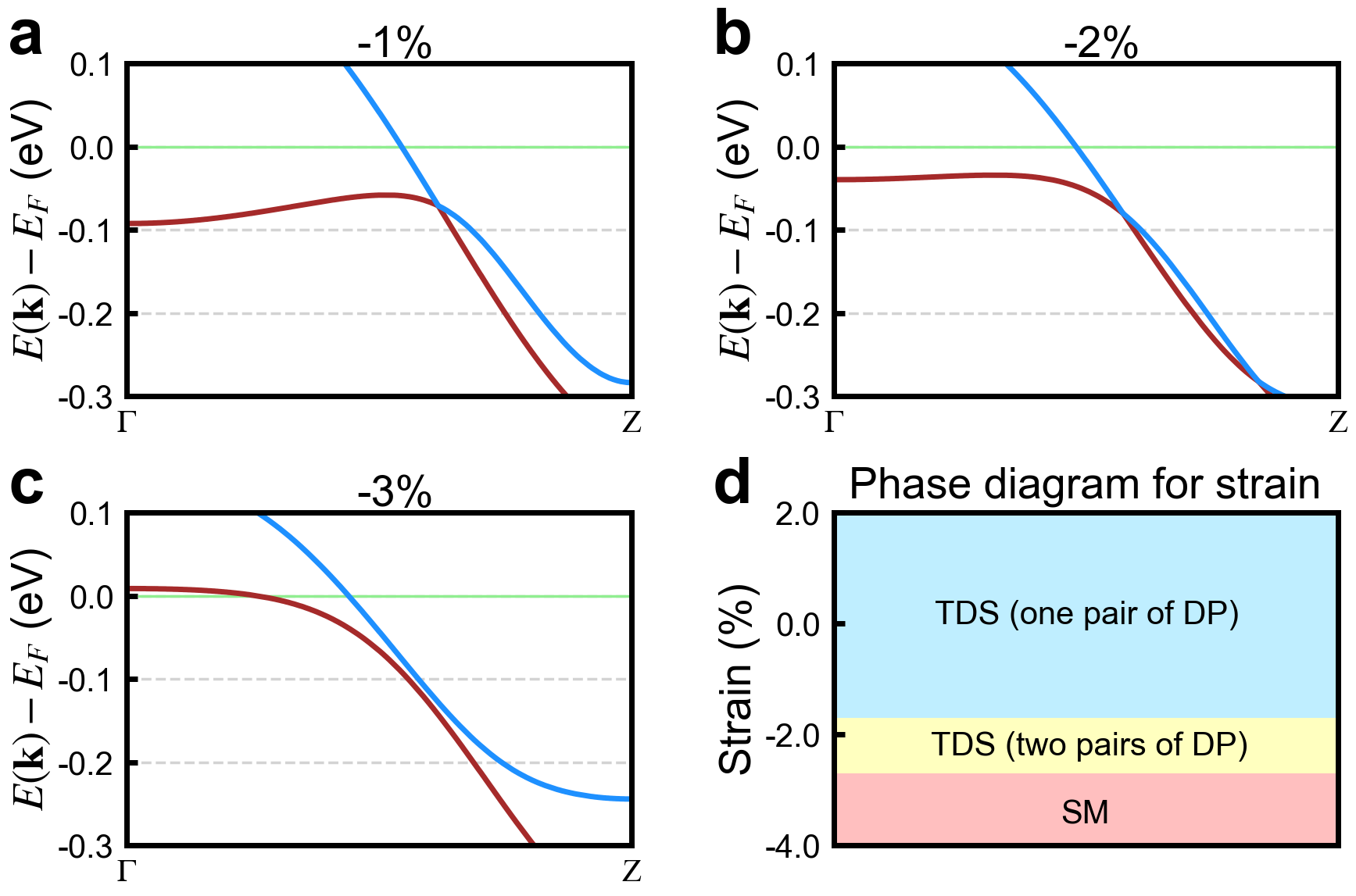}
    \caption{\textbf{Strain effects on the Dirac points of the
        (SrMoO$_{3}$)$_1$/(SrIrO$_{3}$)$_1$ superlattice.} DFT+SOC
      band structure of the (SrMoO$_{3}$)$_1$/(SrIrO$_{3}$)$_1$
      superlattice along the $\Gamma \to \mathrm{Z}$ path under
      different bi-axial strains. Only the lowest conduction band
      (blue curve) and the highest valence band (red curve) are
      shown. The Fermi level is the light green line.  \textbf{a}:
      1$\%$ compressive strain.  \textbf{b}: 2$\%$ compressive
      strain. \textbf{c}: 3$\%$ compressive strain.  \textbf{d}: Phase
      diagram of the (SrMoO$_{3}$)$_1$/(SrIrO$_{3}$)$_1$ superlattice
      as a function of strain. SM means semi-metals and TDS means
      topological Dirac semi-metals.  When the bi-axial strain is from
      2\% tensile to 1.7\% compressive, the superlattice has one pair
      of Dirac points (blue area). When the bi-axial strain is from
      1.7\% compressive to 2.7\% compressive, the superlattice has two
      pairs of Dirac points (yellow area).  When the bi-axial strain
      is from 2.7\% to 4\% compressive, the superlattice does not have
      Dirac points.}
    \label{fig5}
\end{figure}

\clearpage
\newpage

\begin{figure}
    \centering
    \includegraphics[width=0.8\textwidth]{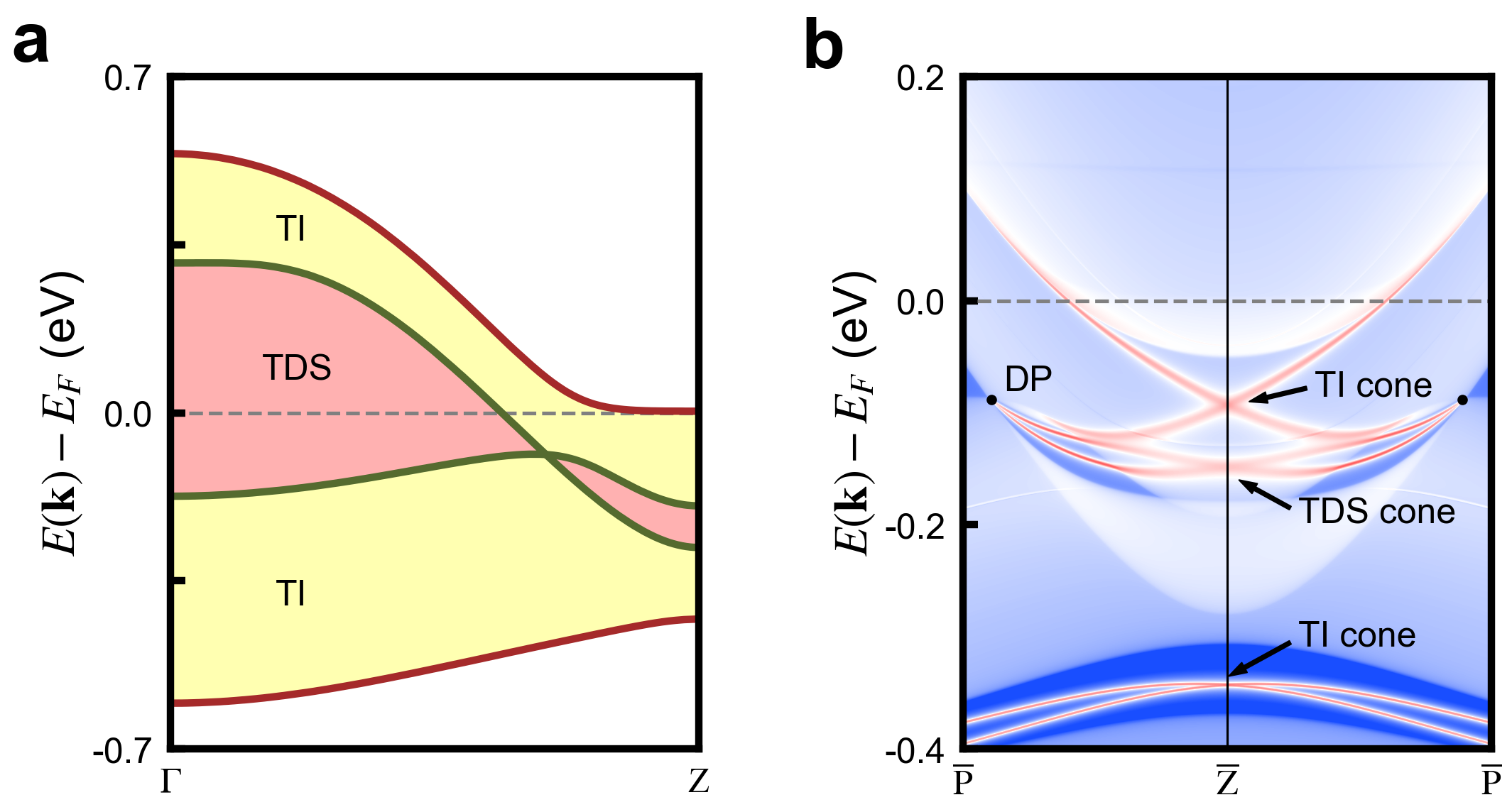}
    \caption{\textbf{Coexisting topological Dirac semi-metal and
        topological insulating states in the
        (SrMoO$_3$)$_1$/(SrIrO$_3$)$_1$ superlattices}. \textbf{a}:
      DFT+SOC band structure of (SrMoO$_3$)$_1$/(SrIrO$_3$)$_1$
      superlattices along $\Gamma$ $\to$ Z path. The two green bands
      are the highest valence band and the lowest conduction
      band. They have a linear crossing along $\Gamma$ to Z with a
      non-trivial mirror Chern number, which leads to a topological
      Dirac semi-metals (TDS) state (highlighted by light red). The
      red bands and green bands do not have crossing in the entire
      Brillouin zone and the gap between them (highlighted by light
      yellow) is topologically nontrivial with $Z_2$ index
      (1;001). \textbf{b}: Band structure of (010) semi-infinite slab
      of the (SrMoO$_3$)$_1$/(SrIrO$_3$)$_1$ superlattice, calculated
      by using the Green-function method. The red indicates the
      surface states. The \textbf{k}-path in the surface Brillouin
      zone is taken to be
      $\mathrm{\overline{P}}\to\mathrm{\overline{Z}}$, where
      $\mathrm{\overline{P}}$ is $(0.0, 0.3)$ and
      $\mathrm{\overline{Z}}$ is $(0.0, 0.5)$. The black dots
      highlight Dirac point (DP). The black arrows highlight TI
      surface Dirac cone and TDS surface Dirac cone. The TDS surface
      Dirac cone is sandwiched between two TI surface Dirac cone in
      the energy-momentum space.}
    \label{fig6}
\end{figure}



\end{document}